\documentclass[prb,aps,twocolumn,showpacs,preprintnumbers,amsmath,amssymb]{revtex4-2}

\usepackage{graphicx}% Include figure files
\usepackage{dcolumn}% Align table columns on decimal point
\usepackage{bm}% bold math
\usepackage{color}
\usepackage{braket}
\usepackage{soul}

\def\EF{$E_\textrm{F}$}

\definecolor{mygray}{gray}{0.6}

\begin{document}

\title{Weyl Fermions with various chiralities in a $f$-electron ferromagnetic system: PrB$_4$}
%\author{Dong-Choon Ryu$^{1,2,3}$, Bongjae Kim$^2$, Chang-Jong Kang$^{3,4}$, B. I. Min$^1$}
%\author{Dong-Choon Ryu$^{1,2,3}$, Junwon Kim$^1$, Kyoo Kim$^4$, Bongjae Kim$^2,6$, Chang-Jong Kang$^{3,5}$, B. I. Min$^1$}
\author{Dong-Choon Ryu$^{1,2,3}$}
\email[dcrhyu@postech.ac.kr]{}
\author{Junwon Kim$^1$}
\author{Kyoo Kim$^4$}
\author{Bongjae Kim$^{2,6}$}
\author{Chang-Jong Kang$^{3,5}$}
\email[Co-corresponding: cjkang87@cnu.ac.kr]{}
\author{B. I. Min$^1$}
\email[Co-corresponding: bimin@postech.ac.kr]{}

\affiliation{
$^1$Department of Physics, Pohang University of Science and Technology, Pohang, 37673, Korea \\
$^2$Department of Physics, Kunsan National University, Gunsan 54150, Korea \\
$^3$Department of Physics, Chungnam National University, Daejeon 34134, Korea \\
$^4$ Korea Atomic Energy Research Institute, Daejeon, Korea  \\
$^5$Institute of Quantum Systems, Chungnam National University, Daejeon, 34134, Korea \\
$^6$Department of Physics, Kyungpook National University, Daegu, 41566, Korea
}
\date{\today}

\begin{abstract}
Rare-earth tetraborides ($R$B$_{4}$) have attracted a lot of recent attention
due to their intriguing electronic, magnetic, and topological properties.
We have theoretically investigated topological properties of PrB$_{4}$,
which is unique among $R$B$_{4}$ family due to its ferromagnetic ground state.
We have discovered that PrB$_{4}$ is an intrinsic magnetic Weyl system
possessing multiple topological band crossings with various chiral charges.
Density-functional-theory band calculations combined with tight-binding band analysis
reveal large Fermi-arc surface states,
which are characteristic fingerprints of Weyl fermions.
Anomalous Hall conductivity is estimated to be very large,
ranging from 500 to 1000 ($\Omega \cdot$cm)$^{-1}$ near the Fermi level,
which also demonstrates the topological Weyl character of ferromagnetic PrB$_{4}$.
These findings suggest that PrB$_{4}$, being a potential candidate of magnetic Weyl system,
would be a promising rare-earth topological system for applications
to next-generation spintronic and photonic devices.
\end{abstract}

\maketitle

%%%%%%%%%%%%%Introduction

\begin{figure}[t]
\includegraphics[width=3.5in]{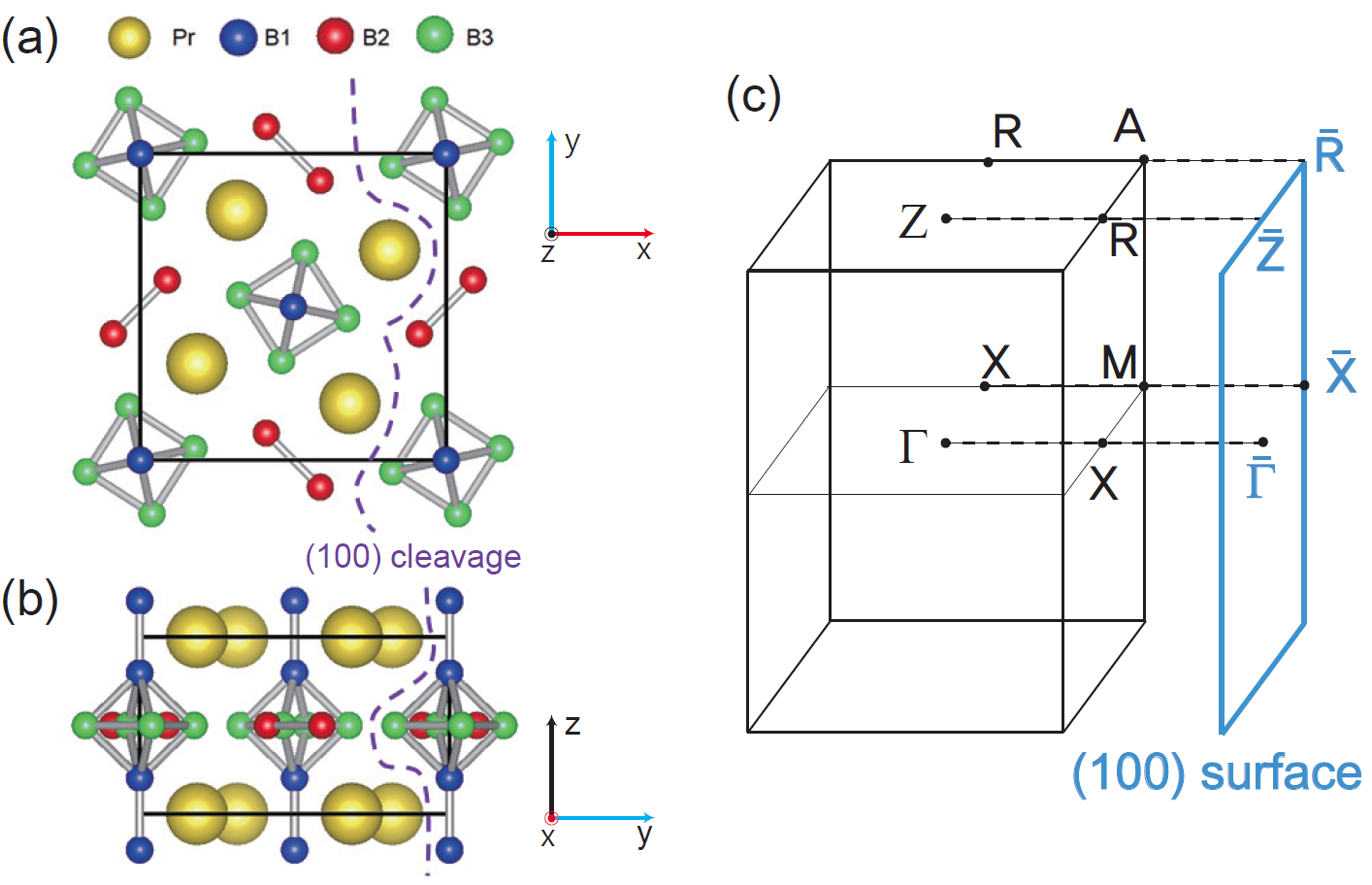}
\caption{(Color Online)
Tetragonal crystal structure of PrB$_4$ with
nonsymmorphic $P4/mbm$ space group.
(a) top view and (b) side view.
Purple dashed lines indicate (100) surface cleavage
we have considered for the Pr-termination in Fig.~\ref{weyl}.
(c) Bulk and (100) surface BZ.
}
\label{cry}
\end{figure}

%--------Fig---------
\begin{figure}[t]
\includegraphics[width=3.5in]{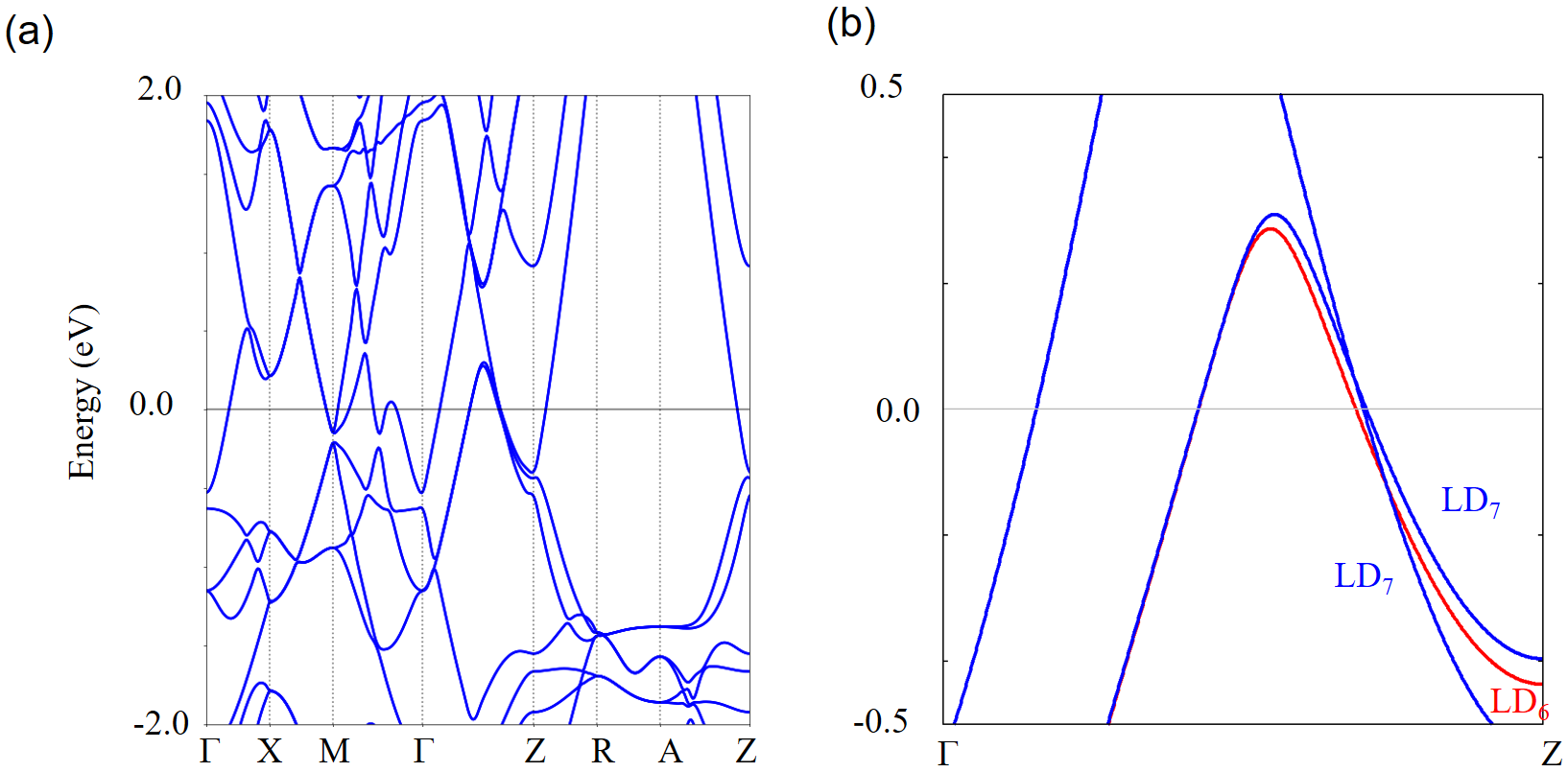}
\caption{(Color Online)
(a) Bulk band structure of nonmagnetic PrB$_4$ obtained
by the 4$f$ opencore calculation,
%{\cblue
including the spin-orbit coupling.
%with spin-orbit coupling.
%}
(b) Amplified band structure near {\EF} along $\Gamma-Z$ with corresponding
IRREP for each band.
The band with the IRREP of LD$_{6}$ is drawn in red for clarity.
}
\label{openbulk}
\end{figure}
%---------fig end ----------------------------------------

\begin{figure*}[t]
\includegraphics[width=7in]{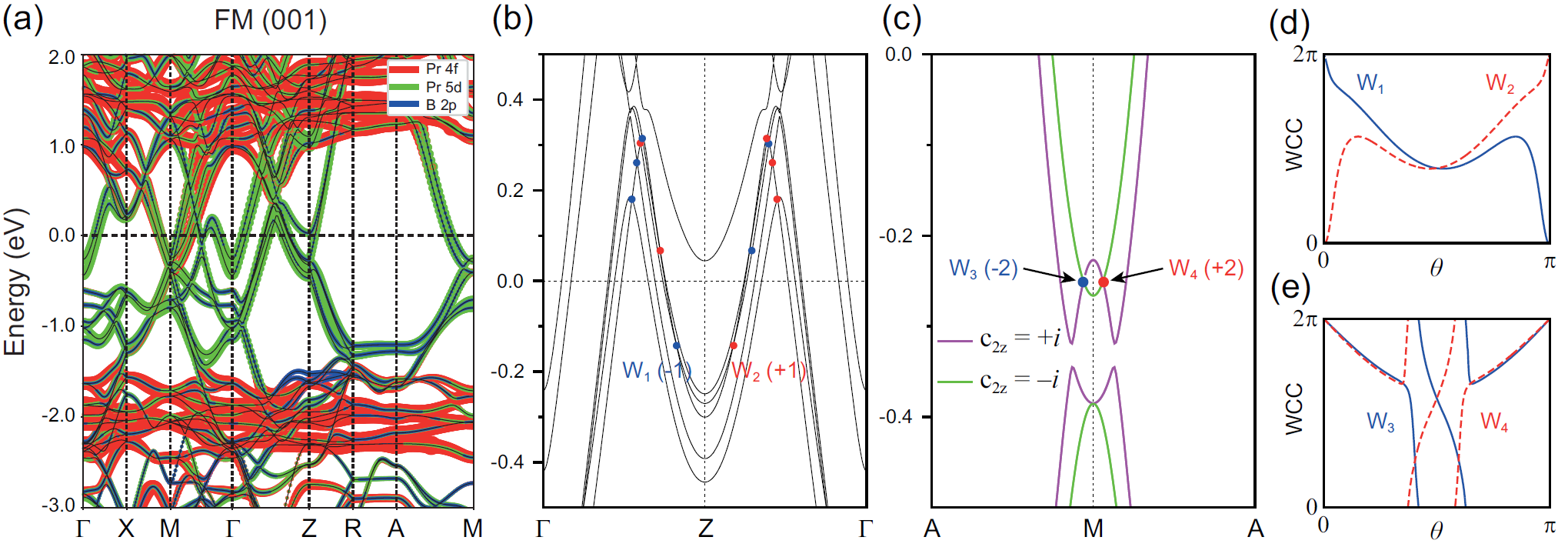}
\caption{(Color Online)
(a) Bulk band structures of PrB$_4$ with FM ordering along the (001) direction.
(b), (c) Bulk band structures along rotationally invariant paths $\Gamma-Z-\Gamma$ and $A-M-A$ under the FM ordering.
The Weyl points of topological band crossings are represented by red and blue dots, depending on their chiral charges.
(d), (e) Evolution of Wannier charge center (WCC) on the spheres centered at four Weyl points $W_1$ - $W_4$ marked in (b) and (c).
}
\label{weyl}
\end{figure*}

\begin{figure*}[t]
\includegraphics[width=7in]{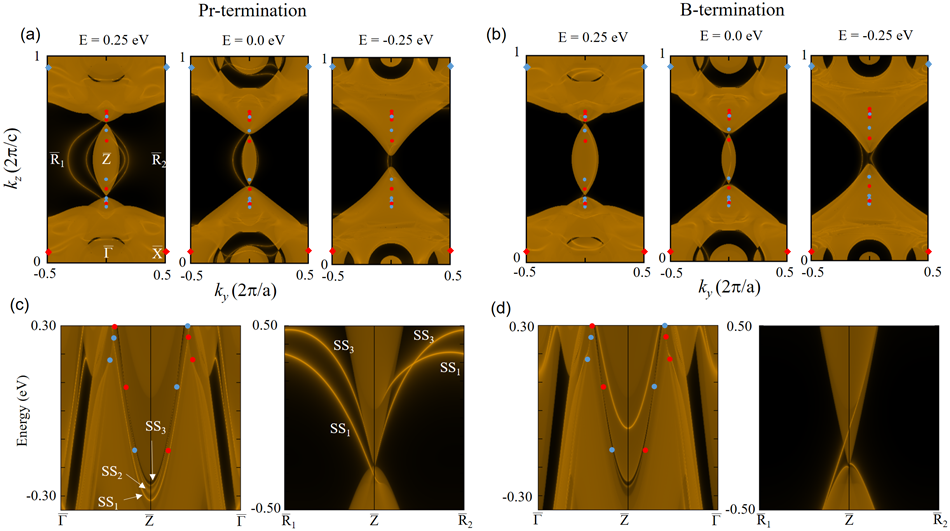}
\caption{(Color Online)
Surface electronic structures on (100) surface of FM PrB$_4$.
(a),(b) Constant-energy surfaces at various energy levels with Pr- and B-termination, respectively.
%{\cblue
Red and blue circles (diamonds) indicate the positions of Weyl fermions with chiralities
$\chi=\pm$1 ($\chi=\pm$2), respectively.
%}
(c),(d) Surface band structures along
$\bar{\Gamma}-\bar{Z}-\bar{\Gamma}$ and
$\bar{R}_1-\bar{Z}-\bar{R}_2$.
Red and blue dots denote the positions of Weyl points
with positive and negative chiral charges, respectively.
}
\label{surf}
\end{figure*}

%--------Fig---------
\begin{figure*}[t]
\includegraphics[width=7in]{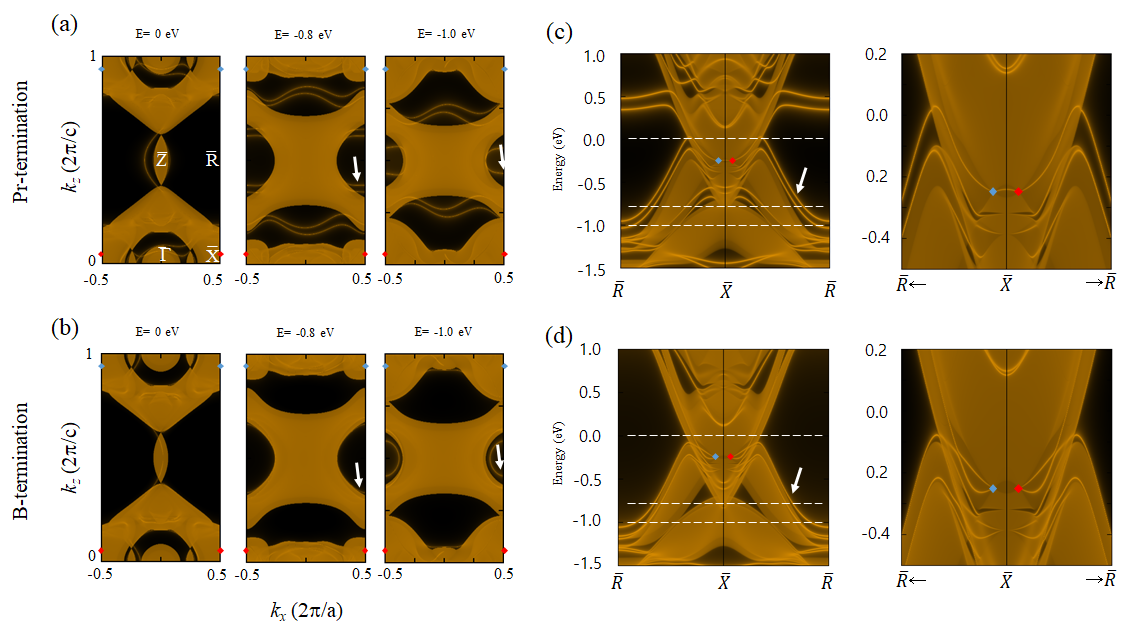}
\caption{(Color Online)
(a), (b) Constant-energy surfaces at various energy levels, and
(c), (d) surface band structures of the ferromagnetic phase of Weyl semimetallic PrB$_4$
for (top) Pr- and (bottom) B-termination.
The rightmost figures correspond to the amplified plots of the left figures in (c) and (d).
The positions of double Weyl fermions
with chirality $|\chi|$ = 2 are denoted by red (positive) and blue (negative) diamonds in the figure.
Surface Fermi-arc states connected to the double Weyl fermions are indicated by white arrows.
The broken lines in (c) and (d) represent the energy levels
of the constant-energy surfaces in (a) and (b).
}
\label{DW}
\end{figure*}
%---------fig end ----------------------------------------

%---------------------------------------

%{\bf Introduction} --
\section{Introduction}

Weyl semimetal is a topological system,
which attracts a lot of recent attention
in the condensed matter physics community \cite{Wan11,Yan17,Armitage18}.
A Weyl semimetal is featured by exotic
bulk Weyl points of twofold-degenerate band crossings
and the associated topological Fermi-arc surface states.
%{\cblue
%With its inherent chiral anomaly,
%This system demonstrates unusual phenomena,
%such as negative longitudinal magnetoresistance,
%chiral magnetic effect, giant anomalous Hall effect \cite{Ong21}.
Due to inherent chiral-anomaly nature of Weyl fermions,
the Weyl system exhibits the negative longitudinal magnetoresistance.
In addition, unusual phenomena of the chiral magnetic effect,
the giant anomalous Hall effect, and the large magneto-optical Kerr effect
are supposed to be realized in the Weyl systems
as a consequence of the chiral anomaly \cite{Yan17,Armitage18,Ong21}.
%}
The emergence of Weyl fermion excitation
requires the breaking of either time-reversal ($T$) or spatial inversion ($P$) symmetry,
because the existence of both symmetries produce the Kramers degenerate bands for all $\bf{k}$,
and thereby any band crossing has fourfold degeneracy.

Since the pioneering study of Weyl semimetal on the $T$-breaking pyrochlore
iridates \cite{Wan11}, subsequent studies of Weyl systems have
focused on $P$-breaking materials,
and so most of reported Weyl systems belong to noncentrosymmetric crystals,
such as transition-metal monophosphides and
dichalcogenides \cite{Weng15, Yang15, Soluyanov15, Sun15}.
In contrast, $T$-breaking magnetic Weyl systems (MWSs) are relatively less explored.
Following earlier theoretical reports on the MWS candidates
of Y$_2$Ir$_2$O$_7$ \cite{Wan11}, HgCr$_2$Se$_4$ \cite{Xu11}, and SrRuO$_3$ \cite{Chen13},
only a few more materials like GdBiPt \cite{Hirschberger16,Shekhar18}, CeSb \cite{Guo17},
and GdB$_4$ \cite{Sohn19} have been proposed as MWS candidates.
These systems, however, demand manipulation of the external magnetic $B$-field to generate Weyl points
or to stabilize the magnetic ordering, and so experimental investigations were limited.
Moreover,
for CeSb, angle-resolved photoemission (ARPES) study raised a question about
the existence of band inversion and the emergence of Weyl fermions \cite{Kuroda18}.

More recently, Heusler-based Co$_2$MnGa \cite{Wang16,Belopolski19}, Kagome-lattice-based Mn$_3$Sn \cite{Kuroda17}
and Co$_3$Sn$_2$S$_2$ \cite{Liu19, Morali19}
were reported to be MWSs.
In the cases of rare-earth systems,
PrAlGe, in which both $P$ and $T$ are broken, was proposed to be an MWS candidate \cite{Chang18,Sanchez20}.
Also, EuB$_6$ was reported to be a candidate for a $T$-breaking MWS \cite{Nie20, Yuan22}.
Hence, there are only a number of genuine $T$-breaking MWS candidates, and they are mostly $d$-electron systems.

In this work, we have investigated topological properties of a
representative rare-earth tetraboride system, PrB$_4$,
and found that PrB$_4$ is a genuine $T$-breaking NWS with $f$-electrons.
With its intrinsic ferromagnetism, PrB$_4$ possesses unique Weyl fermion character with various types of chiral charges.
Note that rare-earth tetraborides $R$B$_4$ ($R$: rare-earth elements)
exhibit diverse magnetic ground states,
depending on $R$ element,
such as Kondo, ferromagnetic (FM), and antiferromagnetic (AFM) states.
Furthermore, exotic topological properties were also predicted
for tetraborides.
As mentioned above, GdB$_4$,
having the in-plane noncollinear AFM ground state of a well-known Shastry-Sutherland lattice type,
was proposed to be a Weyl system in the presence of the external $B$ field \cite{Sohn19}.
Albeit not $R$B$_4$, an actinide-tetraboride Kondo system, PuB$_4$,
was reported to host intriguing fourfold-degenerate topological wallpaper fermions on its surface \cite{Pu}.
Similarly, DyB$_4$, which has the noncollinear AFM ground state,
was reported to host magnetic wallpaper fermions \cite{Dy}.

PrB$_4$ is unique in that it is a sole ferromagnet among $R$B$_4$.
According to a magnetic susceptibility experiment \cite{Wigger05}, upon cooling,
PrB$_4$ shows first the AFM ordering at $T_N \sim$ 19.5 K,
and then the FM ordering below $T_C \sim$ 15.9 K.
In view of the $B$-field-induced MWS for AFM GdB$_4$,
PrB$_4$ in the FM phase is expected to have the Weyl-type band structure
even in the absence of the external $B$-field.
Indeed, for PrB$_4$,
we have theoretically found that Weyl fermions with various chiral charges
emerge in its FM ground state with associated topological Fermi-arc surface states.
Also the estimated anomalous-Hall-conductivity (AHC),
which originates from the large Berry curvature
contributed by some Weyl nodes and
band anticrossings, reaches
as high as 1000 ($\Omega \cdot$cm)$^{-1}$,
which corroborates that PrB$_4$ is a new candidate of intrinsic rare-earth MWS.

%{\bf Crystal structure and computational details} --
\section{Crystal structure and computational details}

PrB$_4$ crystallizes in a tetragonal structure with the nonsymmorphic $P4/mbm$ space group (SG 127).
In Fig.~\ref{cry}, the crystal structure of PrB$_4$ and its bulk and surface Brillouin zone (BZ) are depicted.
The lattice constants and internal coordinates used in this study were adopted from the experiment
($a$=7.235 \AA, $c$=4.116 \AA) \cite{Etou79}.

For the first-principles density functional theory (DFT) band calculations,
we have employed the projector augmented wave (PAW) band method
implemented in VASP in the generalized-gradient approximation (GGA) \cite{PAW,VASP,GGA}.
To describe the strongly-correlated Pr $4f$ electrons,
we have used the GGA+$U$ calculations with Coulomb ($U$) and exchange ($J$) correlation parameters.
We set the parameters for $U$ = 4 eV, which is a commonly accepted value for the Pr atom \cite{Choi09, Destraz20},
and $J$ = 0.4 eV, which reproduces well the observed magnetic moment of 2.1 $\mu_B$ per Pr atom \cite{Wigger05}
(see Fig. S1 in the supplement \cite{Supp}).

Surface electronic structures and chiral charges of Weyl points are obtained
based on a Wannierized tight-binding Hamiltonian \cite{W90}
by utilizing the Wanniertools code \cite{WT}.
We have also obtained the Berry curvature and estimated the anomalous Hall conductivity
based on the Wannierized tight-binding bulk band structures.

%{\bf Bulk band structure of nonmagnetic P\lowercase{r}B$_4$} --
\section{Bulk band structure of nonmagnetic P\lowercase{r}B$_4$}

Figure~\ref{openbulk}(a) shows the bulk band structure of PrB$_4$
obtained by the so-called ``opencore"
calculations, in which Pr 4$f$-electrons are treated as core,
so that the magnetism is suppressed.
In Fig.~\ref{openbulk}(b), the amplified band structure near the Fermi level ({\EF})
along $\Gamma-Z$ is plotted with the irreducible representation (IRREP) of each band.
%IRREPs of LD$_6$ and LD$_7$ indicate that all the bands along $\Gamma-Z$ are twofold-degenerate,
%and so the fourfold-degenerate band crossing can occur when two bands of distinct IRREPs intersect.
%Since the system preserves both the inversion $P$ and time-reversal $T$ symmetries
%due to the absence of magnetism,
%any fourfold band crossing in this system must be a Dirac point.
Every band along $\Gamma-Z$ is twofold-degenerate due to the time-reversal pairing,
and so the fourfold-degenerate band crossing can occur when two bands of distinct IRREPs of LD$_6$ and LD$_7$ intersect,
as shown in Fig.~\ref{openbulk}(b).
Since the system preserves both the inversion $P$ and time-reversal $T$ symmetries
in the absence of magnetism,
any fourfold band crossing in this system must manifest itself as a Dirac point.
Note that the Dirac point shown in Fig.~\ref{openbulk}(b) is of type-II nature.
Surface states relevant to those Dirac points are further discussed
in the supplement (Fig. S3) \cite{Supp}.

%{\bf Weyl points and Fermi arcs} --
\section{Weyl points and Fermi arcs}

Now we discuss the role of the magnetism in the emergence of the Weyl fermions in PrB$_4$.
First, we have investigated energetics of three different magnetic structures of PrB$_4$,
FM and two-types of AFM,
and found that the FM state is indeed a ground state with magnetic ordering
along the (001) direction, which is in agreement with the experimental
results (see Fig. S2 and Table S1 in the supplement \cite{Supp}).
Then the formation of Weyl points, when breaking the time-reversal symmetry,
is examined on the basis of the model Hamiltonian obtained from the nonmagnetic calculation
by taking into account a Zeeman-like term
(see Fig. S5 and S6 in the supplement \cite{Supp}).

In order to examine the existence of Weyl points and their associated Fermi-arc surface states,
we have investigated the band structures of FM PrB$_4$
with (001) magnetic ordering.
Figure~\ref{weyl}(a) shows that
Pr $4f$ states are located near $-2$ eV and 2 eV, while
wider Pr $5d$ and B $p$ bands are located near the Fermi level {\EF}.

The high symmetry $k$-paths, namely $\Gamma-Z$ and $M-A$,
are invariant under $C_{4z}$ and $C_{2z}$ rotations, as shown in Fig.~\ref{cry},
which leads to the occurrence of band crossings along those two paths.
Indeed, the band crossings in the vicinity of {\EF}
are clearly manifested
in Figs.~\ref{weyl}(b) and (c) along $\Gamma-Z-\Gamma$ and $A-M-A$ paths, respectively.
The red and blue dots represent the Weyl points with positive and negative chiralities, respectively,
which are identified from the Wilson loop calculations.
There are several Weyl points along $\Gamma-Z$, while there is only one
along $A-M$ near {\EF} (see Table S2 in the supplement \cite{Supp})

Notable in Figs.~\ref{weyl}(b) and (c) is that Weyl points on the $\Gamma$-$Z$ path
are conventional twofold-degenerate Weyl nodes,
but those on the $M-A$ path, $W_3$ and $W_4$, are exotic fourfold-degenerate Weyl nodes,
arising from the unique crystal symmetry of FM PrB$_4$.
Because of the magnetic ordering, the $T$ symmetry is not preserved here, but
the combination of $T$ and nonsymmorphic screw-axis symmetry,
$S_y$ = $\{C_{2y}\vert\frac{1}{2}\frac{1}{2}0\}$, is preserved.
Interestingly, this combined symmetry $S_yT$ makes every band on the $M-A$ path
doubly degenerate even under the FM ordering.
To be more specific,
for any eigenstate $\psi$ on $M$-$A$,
there exists its Kramers pair $S_yT\psi$ since $(S_yT)^2 = \exp(ik_x) = -1$ on $M$-$A$
%($(k_x,k_y)$ = $(\pi,\pi)$).
(Note that both $M$ and $A$ have $(k_x,k_y)$ = $(\pi,\pi)$).
Furthermore, when $C_{2z}\psi$ = $\pm i\psi$, we have $C_{2z}(S_yT\psi) = -\exp[i(k_x + k_y)]S_yTC_{2z}\psi = \pm i(S_yT\psi)$.
Namely, two bands in each Kramers pair have the same $C_{2z}$ eigenvalues,
which suggests that some accidental band crossings between two Kramers pairs
with distinct $C_{2z}$ eigenvalues can be preserved by crystal symmetry.
In fact, the Wannier charge center (WCC) calculations in Figs.~\ref{weyl}(d) and (e) confirmed
that the chiralities (topological charges: $\chi$'s) of ($W_1$ and $W_2$) and ($W_3$ and $W_4$)
Weyl points are $\chi=\pm$1 and $\chi=\pm$2,
implying the single-Weyl and double-Weyl points, respectively.
The existence of both single- and double-Weyl points in PrB$_4$, albeit somewhat complicated,
would be more effective for applications to topological devices,
as in the case of chiral fermion systems with multifold degeneracy \cite{Chang17,Hasan22}.

To confirm the Fermi arcs, which are one of the hallmarks of Weyl fermions,
the surface electronic structure calculations were carried out.
Although the (001) surface is the natural cleavage plane of PrB$_4$,
the (100) surface is more preferable
to identify the Fermi arcs more clearly.
On the (001) surface, all the Weyl points on $\Gamma-Z$ and $M-A$ are to be
projected onto $\bar{\Gamma}$ and $\bar{M}$, respectively.
This results in the overlap of Weyl points of opposite chiralities
and, as a consequence, no vestige of the Fermi arcs.
So we have examined the surface electronic structures on the (100) surface.

Figure~\ref{surf} shows the two different sets of possible Fermi arcs
depending on the surface terminations.
In Fig.~\ref{surf}(a), which displays constant-energy surfaces
for the Pr-termination, a few surface states presumed to be Fermi arcs
are observed near $\bar{Z}$,
and those Fermi arcs become larger for higher energy cut.
In Fig.~\ref{surf}(c) are plotted surface band structures along
$\bar{\Gamma}-\bar{Z}-\bar{\Gamma}$ and
$\bar{R}-\bar{Z}-\bar{R}$, in which
three surface states SS$_1$ - SS$_3$ are identified.
Comparative analysis of Fig.~\ref{surf}(a) and (c) suggests that
SS$_1$ corresponds to the largest Fermi arc shown in Fig.~\ref{surf}(a)
connecting two charge-opposite Weyl points at higher energy, slightly below $E = 0.3$ eV,
while SS$_2$ and SS$_3$ connect four Weyl points
close to $\bar{Z}$.
On the other hand, for the B-termination of Fig.~\ref{surf}(b),
putative Fermi arcs are buried in bulk Fermi surface near $\bar{Z}$
and so are difficult to be discerned.
Nevertheless, Figs.~\ref{surf}(b) and (d) reveal that
there is a narrow band gap at $\bar{Z}$ near $E = -0.25$ eV,
and Fermi arcs can be resolved in the gap region.

In contrast, the Fermi arcs arising from the Double Weyl points are difficult to identify,
because they are mostly buried within the bulk continuum.
Nevertheless, surface band structure and constant-energy surface calculations in Fig.~\ref{DW}
show that one of the long tails connected to those double Weyl points (indicated by white arrows in Fig.~\ref{DW}(c),(d))
is partially revealed between $\bar{X}$-$\bar{R}$. Indeed, the zoomed-in figures in (c) and (d) (the rightmost figures)
of Fig.~\ref{DW} clearly show the emergence of associated surface states from the double Weyl points,
despite being buried within the bulk continuum.

It is worthwhile to compare the topological properties of PrB$_4$
depending on its different magnetic states.
As discussed earlier, we have shown that nonmagnetic PrB$_4$ is a Dirac semimetal,
hosting a type-II Dirac point along $\Gamma-Z$.
%{\cred
%For AFM PrB$_4$, we have found that it can be just a trivial normal metal or MWS, as discussed
%in Fig. S4 of the supplement \cite{Supp}.
%However, the possible AFM MWS state
%does not host a Weyl point on $\Gamma-Z$
%nor a double-Weyl point on $M-A$,
%while it host Weyl points on generic k-points
%near M.
%Therefore, it is expected that PrB$_4$, upon cooling, undergoes the topological transitions
%from a Dirac to a trivial or , and then to a Weyl system,
%as the system transforms from PM to AFM and then to FM phase.
For AFM PrB$_4$, we have found that it can be either a trivial normal metal (AFM1 configuration)
or MWS (AFM2 configuration) depending on magnetic configurations,
as demonstrated in Figs. S2 and S4 of the supplement~\cite{Supp}.
However, the AFM2 MWS state does not host a Weyl point on $\Gamma-Z$
nor does a double-Weyl point on $M-A$.
Instead, it hosts Weyl points on generic $k$-points near $M$
%{\cblue
(see Table S3 in the supplement~\cite{Supp}).
%}
Notably, PrB$_4$ exhibits a unique temperature-dependent variation in magnetic ordering upon cooling,
transitioning from PM to AFM and finally to FM states.
This leads to a topological phase transition from a topological Dirac semimetal to an intermediate AFM metal and,
ultimately, to a magnetic Weyl semimetal as the temperature decreases.
This property allows for the utilization of temperature-tuned topological properties in PrB$_4$.
%}

%-----------------------------
\begin{figure}[t]
\includegraphics[width=3.5in]{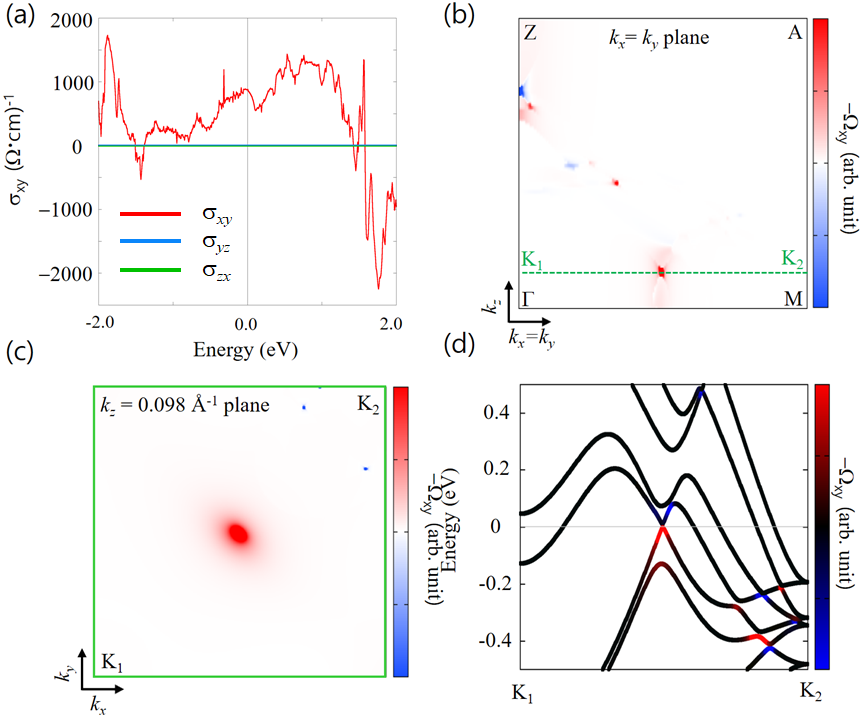}
\caption{(Color Online)
(a) Chemical-potential energy dependent $\sigma_{xy}$ for FM PrB$_4$.
(b), (c) Berry curvature $\Omega_{xy}(\boldsymbol{k})$ plot on $k_x = k_y$ and $k_z$ = 0.098$\AA^{-1}$ planes at {\EF}, respectively.
$K_1$ and $K_2$ are $\boldsymbol{k}$-points on the $k_z$ = 0.098$\AA^{-1}$ plane.
(d) Band-resolved contribution to $\Omega_{xy}(\boldsymbol{k})$ along the $K_1-K_2$  path.
}
\label{ahc}
\end{figure}
%-----------------------------

%{\bf Anomalous Hall conductivity (AHC)} --
\section{Anomalous Hall conductivity (AHC)}

In MWSs, the AHC, which is much larger than the ordinary Hall conductivity,
originates from intrinsic topological properties of the band structure,
specifically the Berry curvature.
We have estimated the AHC, $\sigma_{\alpha\beta}$, using the following equations \cite{AHC}:
\begin{equation} \label{eq1}
\sigma_{\alpha\beta} = -\frac{e^2}{\hbar}\int_{BZ}
\frac{d\boldsymbol{k}}{(2\pi)^3}\Omega_{\alpha\beta}(\boldsymbol{k}),
\end{equation}
where $\Omega_{\alpha\beta}(\boldsymbol{k})$ is the total Berry curvature,

\begin{equation} \label{eq2}
\Omega_{\alpha\beta}(\boldsymbol{k}) = -2 Im \sum_{v}\sum_{c}
\frac{
v_{vc,\alpha}(\boldsymbol{k})v_{cv,\beta}(\boldsymbol{k})
}{
[\epsilon_c(\boldsymbol{k})-\epsilon_v(\boldsymbol{k})]^2.
}
\end{equation}
Here $\epsilon_n(\boldsymbol{k})$ is the energy of $n$-th band at $\boldsymbol{k}$,
$c$ and $v$ denote unoccupied and occupied bands,
and the velocity $v_{nm,\alpha}(\boldsymbol{k})$ is given by
\begin{equation} \label{eq3}
v_{nm,\alpha}(\boldsymbol{k})
= \langle\psi_{n\boldsymbol{k}}
| \hat{v}_{\alpha} |
\psi_{m\boldsymbol{k}}\rangle
= \frac{1}{\hslash}
\langle u_{n\boldsymbol{k}}
|
\frac
{\partial{\hat{H}}(\boldsymbol{k})}
{\partial{k_\alpha}}
|
u_{m\boldsymbol{k}}\rangle.
\end{equation}
For FM state with magnetic ordering along (001) direction,
the mirror symmetry $M_z$ is present, which prohibits the $\sigma_{yz}$ or $\sigma_{zx}$ component,
resulting in only $\sigma_{xy}$ being finite.

In Fig.~\ref{ahc}(a), we have shown the chemical-potential energy dependent AHC, $\sigma_{xy}(E)$, for FM PrB$_4$.
The obtained $\sigma_{xy}(E)$ is highly
nonmonotonic with respect to energy position, and is very large
ranging from 500 $\Omega^{-1}$cm$^{-1}$
to 1000 $\Omega^{-1}$cm$^{-1}$ near {\EF} $(E=0)$.
The peaks and dips in the energy dependence of $\sigma_{xy}(E)$ are expected to arise from
the large Berry curvature at the corresponding energy.
To explore the origin of such large AHC at {\EF}, we have examined Berry curvature
$\Omega_{xy}(\boldsymbol{k})$
on the two planes, $k_x = k_y$ and  $k_z = 0.098\AA^{-1}$.
As plotted in Fig. \ref{ahc}(b) and (c),
there are multiple sources of Berry curvature.
Among them, those near $\Gamma-Z$ path in Fig. \ref{ahc}(b) are related to the Weyl points.
According to Eq.~(\ref{eq2}),
each Weyl point gives rise to huge AHC contribution, since
$\epsilon_c(\boldsymbol{k})-\epsilon_v(\boldsymbol{k})$
goes to zero.
However,
since Weyl points are pair-created with opposite chirality,
the sum of their Berry curvatures over the whole BZ would vanish.

Noteworthy is that
large $\Omega_{xy}(\boldsymbol{k})$ is observed
near the center of $\Gamma$-$M$,
which becomes the largest near $k_z = 0.098\AA^{-1}$ (Fig. \ref{ahc}(c))
would yield the large AHC.
Indeed, band-resolved contribution to $\Omega_{xy}(\boldsymbol{k})$
along $K_{1}-K_{2}$ in Fig.~\ref{ahc}(d)
clearly indicates that
the large Berry curvature and AHC originate from
the crossing-like band dispersion located inbetween $K_{1}-K_{2}$ near {\EF}.
%In fact, there exists here a slight gap with the size of 10 meV,
%and so the band dispersion has an anticrossing feature.
%Since this contribution is not from Weyl point,
Significantly, a 10 meV gap is present there,
implying an anticrossing feature in the band dispersion.
Since this contribution does not arise from Weyl point,
and  any crystal symmetry $g \in \{P, M_z, C_{4z}\}$ in PrB$_4$
guarantees $\Omega_{xy}(\boldsymbol{k}) = \Omega_{xy}(g\boldsymbol{k})$,
there is no cancellation
in the total Berry curvature.

%{\bf Concluding Remarks} --
\section{Concluding Remarks}

We predict that a representative FM rare-earth tetraboride, PrB4, hosts the
multiple Weyl fermions with various topological charges,
namely fourfold degenerate double-Weyl point with charge $\pm$2
as well as conventional twofold degenerate Weyl point with charge $\pm$1.
Such a multitude of topological charges would facilitate PrB$_4$ to be more advantageous
than other MWSs having just conventional single-Weyl nodes,
because one can explore
the stronger topological effects,
easier manipulability, and more stability,
as in the case of the chiral fermion systems with multifold degeneracy.
Furthermore, due to its unique temperature-dependent variation of magnetic ordering upon cooling,
from paramagnetic (PM), AFM to FM,
one can make use of temperature-tuned topological properties of PrB$_4$.
Further experimental studies are encouraged to validate our theoretical predictions
and extend our understanding of this fascinating material.
%{\cred
%Also, our results can provide information on the study for topological properties of
%ErB$_4$ and TmB$_4$,
%since AFM structures of PrB$_4$ we investigated are same as those of
%ErB$_4$ and TmB$_4$ \cite{Ye17}.
Furthermore, our findings hold implications for exploring
the topological properties of ErB$_4$ and TmB$_4$
since the AFM structures of PrB$_4$ we investigated in this study are identical to
those of ErB$_4$ and TmB$_4$ \cite{Ye17}. \\ \\
%}

%\textit{Acknowledgment}---
{\bf Acknowledgments}

Helpful discussions with J.-S. Kang are greatly appreciated.
D.-C. R. and C.-J. K. were supported by NRF (Grant No. 2022R1C1C1008200) and
the KISTI Supercomputing Center (Project No. KSC-2024-CRE-0050).
D.-C. R. was supported by NRF (Grant No. RS-2023-00274550).
C.-J. K. was also supported by the National Research Foundation of Korea Grant funded by
the Korean Government (MOE).
K. K. was supported by the internal R\&D program at KAERI (No.524550-24).
B.K. acknowledges support by NRF Grants (No. 2021R1C1C1007017 and No. 2022M3H4A1A04074153)
and KISTI supercomputing Center (Project No. KSC-2022-CRE-0465).

\end{document}